\documentclass[twocolumn,aps,showpacs,prb,tightenlines,amsmath,amssymb]{revtex4}
\usepackage{graphicx}
\usepackage{amssymb}
\usepackage{dcolumn} 
\usepackage{amsmath}
\usepackage{bm}% bold math
\usepackage{colordvi}
\newcommand{\bgreek}[1]{\mbox{\boldmath$#1$\unboldmath}}
\makeatletter

\newcommand{\Rmnum}[1]{\expandafter\@slowromancap\romannumeral #1@}
\makeatother

\begin{document}

\title{Spin diffusion in $n$-type (111) GaAs quantum wells}
\author{B. Y. Sun} 
\author{K. Shen}
\thanks{Author to  whom correspondence should be addressed}
\email{kkshen@mail.ustc.edu.cn.}
\affiliation{Hefei National Laboratory for Physical Sciences at
  Microscale and Department of Physics,
University of Science and Technology of China, Hefei,
  Anhui, 230026, China}
\date{\today}

\begin{abstract}
We utilize the kinetic spin Bloch equation approach to investigate the
steady-state spin
diffusion in $n$-type (111) GaAs quantum wells, where the in-plane
components of the Dresselhaus spin-orbit 
coupling term and the Rashba term can be partially canceled by each other. A
peak of the spin diffusion length due to the cancellation is 
predicted in the perpendicular electric field dependence. It is shown that
the spin diffusion length around the peak can be markedly controlled via
temperature and doping. When the electron gas enters into the degenerate regime,
the electron density also leads to observable influence on the spin
diffusion in the strong cancellation regime.
Furthermore, we find that the spin diffusion always presents strong anisotropy
with respect to the direction of the injected spin polarization.
The anisotropic spin diffusion depends on whether the electric field is far away
from or in the strong cancellation regime. 
\end{abstract}

%71.10.-w       Theories and models of many-electron systems;
%71.70.Ej       Spin-orbit coupling, Zeeman and Stark splitting, Jahn-Teller effect
%72.25.Dc	Spin polarized transport in semiconductors
%73.63.Hs	Quantum wells

\maketitle
\section{INTRODUCTION}
The spintronics with the aim to incorporate the spin freedom of the
electrons to the traditional electronic devices to design novel
devices has attracted much
attention in the past decades.\cite{opticalorientation,Awschalom,Zutic,Glazov,Chen,wuReview,Korn,Wolf} 
For the application of such spintronic devices, the suitable spin relaxation
time and spin diffusion/injection length are of essential
importance,\cite{Tombros,DattaDas,Wunderlich,Koo} which requires
comprehensive investigation for
the thorough understanding of the spin properties in these systems.
Actually, the spin relaxation and spin diffusion/injection have been widely
investigated both experimentally and theoretically for a long period.
\cite{opticalorientation,wuReview,Glazov,Zutic,Chen,Awschalom,Korn,Wolf}

In $n$-type III-V zinc-blende semiconductors, the D'yakonov-Perel' (DP) 
mechanism,\cite{Dyakonov} which results from the spin
precession under the momentum-dependent effective magnetic field
(inhomogeneous broadening\cite{wu}) together with any scattering
process, is identified as the predominant spin relaxation 
mechanism.\cite{Awschalom,wuReview,jiang,jiang2}
In the absence of the external magnetic field, the inhomogeneous
broadening is mainly supplied by the spin-orbit couplings composed of
the Dresselhaus term\cite{Dresselhaus} and the Rashba
one.\cite{Rashba} While the former one is due to the bulk inversion asymmetry of
the crystal, the latter one originates from the structure inversion asymmetry
and is tunable, e.g., via the electric field along the growth direction.
The spin manipulation based on the competition of the Dresselhaus
and Rashba spin orbit couplings is an interesting issue in semiconductor
spintronics.\cite{Averkiev,Schliemann,Cheng2,Zarbo,Sun,Cartoixa,Vurgaftman}
In the previous works, it was shown that the electron spin relaxation time in
$n$-type (111) GaAs quantum wells (QWs) can be significantly enhanced when the
in-plane components of the Dresselhaus term in the vicinity of the Fermi surface
are strongly canceled by the Rashba term.\cite{Sun, Cartoixa, Vurgaftman} 
This suggests the intriguing spin diffusion property in this system,
which is however still not very clear, to our best
knowledge. Therefore, the goal of this paper is to supply
more knowledge for the spin diffusion in (111) GaAs QWs.

Differing from the spin relaxation in time domain, the inhomogeneous broadening
in spin diffusion is determined by ${\bgreek\omega_{\bf k}}=m^\ast[{\bf
    \Omega}({\bf k})+g\mu_B{\bf B}]/k_x$,\cite{Cheng} where ${\bf \Omega}$ and
${\bf B}$ are the spin-orbit coupling and external magnetic field,
respectively. According to the previous works, the spin diffusion
properties are strongly dependent on the detailed form of the
inhomogeneous broadening. Specifically, in $n$-type (001) GaAs
QWs where the inhomogeneous broadening is introduced by Dresselhaus spin-orbit
coupling, Cheng {\em et al.}\cite{Cheng} showed that the spin diffusion
is enhanced by the scattering but suppressed with the increase
  of the temperature in the strong scattering regime. In
contrast, the spin diffusion length is reduced by increasing the
scattering strength when the scattering is rather weak.
For the inhomogeneous broadening supplied solely by the external
magnetic field in Si/SiGe QWs, Zhang and Wu\cite{Zhang}
found that the spin diffusion is suppressed by scattering
monotonically. Interestingly, the spin diffusion length is reduced
by the scattering in the weak scattering regime and turns to be a constant in
the strong scattering limit (independent of the electron density, temperature,
and scattering strength) in the monolayer graphene, where the Rashba spin-orbit 
coupling is dominant.\cite{Zhanggraphene} 
Due to the partial cancellation between the Dresselhaus and Rashba terms,
the influence of the electron density,
temperature, and scattering strength on the spin diffusion
length in $n$-type (111) GaAs QWs is expected to be an interesting issue.

In the present paper, we investigate the steady-state spin diffusion in
$n$-type (111) GaAs QWs by solving the kinetic spin Bloch equations
(KSBEs),\cite{wuReview} which include all the relevant scatterings,
i.e., electron-impurity, 
electron-acoustic/longitudinal-optical-phonon, and
electron-electron scatterings. This work only focuses on the DP-limited spin
relaxation in the strong
scattering regime. We find that the electric field (along
the growth direction) dependence of the spin diffusion length shows a peak
due to the strong cancellation between the Dresselhaus and Rashba spin-orbit
couplings. 
We find that the spin diffusion length, out of the strong cancellation
regime, is insensitive to the
temperature, doping density and electron 
density. However, the spin diffusion length is markedly manipulated by
changing temperature and doping density around the peak. For the degenerate
electron gas, the electron density can also lead to observable influence on the
spin diffusion length.
Last but not the least, we investigate the anisotropy of the spin diffusion with
respect to the spin polarization direction of the injected electrons.
The analytical solution of the KSBEs with only electron-impurity scattering 
is also presented to explain the numerical calculation.

This paper is organized as follows. In Sec.\,{\Rmnum 2}, we
introduce the KSBEs and give an analytical investigation for the case with
only the electron-impurity scattering. In Sec.\,{\Rmnum 3}, the spin
diffusion is investigated by solving the KSBEs numerically with all
the relevant scatterings included. A brief summary is given
in Sec.\,{\Rmnum 4}.

\section{KSBEs AND THE ANALYTICAL INVESTIGATION}

We start our investigation from $n$-type (111) GaAs
QWs under the infinite-depth-square-well
approximation. The well width 
$a$ is taken to be 7.5~nm and only the lowest subband is relevant in our 
investigation. The electrons with their spins polarized along the direction
${\bf \hat n}$ are injected 
into the QWs at the left boundary ($x=0$) and diffuse along the
$x$-axis. The right boundary is set to be $x=L$ with $L$ much longer
than the 
spin diffusion distance. Therefore, the spin polarization of the electrons at
the right boundary is always set to be zero. Since no magnetic field is applied in our
model, the spinors precess only due to the DP term. Setting the $z$-axis
along the growth direction [111], $x$-axis along
[1$\bar{1}$0] and $y$-axis along [11$\bar{2}$], the total DP term
reads 
\begin{eqnarray}
  \begin{cases}
    \Omega_{x}({\bf k})=\gamma({-k^2+4{\langle k_z^2\rangle}})k_y/{2\sqrt{3}} -\alpha
    eE_zk_y,\\
    \Omega_{y}({\bf k})=-\gamma({-k^2+4{\langle k_z^2\rangle}})k_x/{2\sqrt{3}}+\alpha
    eE_zk_x,\\
    \Omega_{z}({\bf k})=\gamma({k_x^3-3k_xk_y^2})/\sqrt{6}.
  \end{cases}
  \label{eq1}
\end{eqnarray}
Here, $\gamma$ and $\alpha$ are 
the Dresselhaus and Rashba spin-orbit coupling coefficients,
respectively. $E_z$
stands for the electric field along $z$-axis and ${\langle
  k_z^2\rangle}=(\pi/a)^2$ is the average of the operator $-(\partial/\partial 
z)^2$ over the electron state of the lowest subband.
It is clear that for the particular electric field
\begin{equation}
  E^c_z=(4{\langle k^2_z \rangle}-\overline{k^2})\gamma/(2\sqrt{3}\alpha e),
  \label{eq2}
\end{equation}
the in-plane components of the Dresselhaus spin-orbit coupling are strongly canceled
by the Rashba term.\cite{Cartoixa,Vurgaftman,Sun} 
$\overline{k^2}$ here is the average of $k^2$ over the imbalance of the
  spin-up and -down electrons.

The KSBEs read\cite{wuReview,wu}
\begin{eqnarray}
&&\hspace {-0.4 cm}
  \partial_t{\rho_{\bf k}(x,t)}=-{e}{\partial_x \Psi(x,t)}{\partial_{k_x} \rho_{\bf k}(x,t)}-({k_x}/{m^\ast}){\partial_x \rho_{\bf k}(x,t)}  \nonumber \\
  &&\hspace{0.8 cm}\mbox{}+{\partial_t\rho_{\bf
      k}(x,t)}\big|_{\mbox{coh}}+{\partial_t\rho_{\bf
      k}(x,t)}\big|_{\mbox{scat}}.
  \label{eq3}
\end{eqnarray}
Here, $\rho_{\bf k}(x,t)$ are the single-particle density matrices of electrons
with the in-plane wave-vector ${\bf k}$ at position $x$ and time $t$.
  $m^\ast$ is the effective mass.
$\Psi(x,t)$ is the electric potential satisfying the Poisson equation
$\nabla^2_x\Psi(x,t)=e[N_e(x,t)-N_0]/(a 
\kappa_0 \varepsilon_0)$ with $N_e(x,t)=\sum_{\bf k}{\rm Tr}[\rho_{\bf k}(x,t)]$
standing for the local electron density. $N_0$
is the background positive charge density and $N_e(x,0)=N_0$ denoting the initial
  condition. $\varepsilon_0$ and 
$\kappa_0$ are the vacuum and relative static dielectric constants,
respectively. ${\partial_t\rho_{\bf 
    k}(x,t)}\big|_{\mbox{coh}}=-{i}[{\bgreek \Omega}({\bf 
    k})\cdot{\bgreek \sigma}/2+\Sigma_{\bf k}(x,t),\rho_{\bf k}(x,t)]$ is the coherent term
  with $\Sigma_{\bf
    k}(x,t)=-\sum_{\bf q}V_q\rho_{{\bf k}-{\bf q}}(x,t)$ being the Hartree-Fock 
  term.\cite{Haug} Here, ${\bgreek \sigma}$
  are the Pauli matrices and $V_q$ is the Coulomb potential 
  within the random phase approximation.\cite{Zhou} The scattering term
${\partial_t \rho_{\bf 
    k}(x,t)}\big|_{\mbox{scat}}$ includes the electron-impurity,
electron-acoustic/longitudinal-optical-phonon, and electron-electron
scatterings, of which the expressions can be found in
Refs.~\onlinecite{Zhou} and \onlinecite{weng}.

To speculate the properties of the spin diffusion, we first simplify
the KSBEs with only the elastic impurity scattering included in the scattering term.
Then, by taking the steady-state condition, i.e., $\partial_t{\rho_{\bf k}(x,t)}=0$,
and performing the Fourier transformation with respect to $\theta_{\bf k}$ 
[with ${\bf k}=k(\cos\theta_{\bf 
  k},\sin\theta_{\bf k})$], one obtains
\begin{eqnarray}
  \nonumber
  \partial_x(\rho^{l+1}_k+\rho^{l-1}_k)&=&-{2}{\rho^l_k}{/({v_k}\tau^l_k)}-\gamma_k[\sigma_z,\rho^{l-3}_k+\rho^{l+3}_k]\\
  &&\hspace{-0.5cm}\mbox{}-\beta_k[\sigma_+,\rho^{l+1}_k]+\beta_k[\sigma_-,\rho^{l-1}_k],
  \label{eq4}
\end{eqnarray}
where $\sigma_{\pm}=(\sigma_x\pm i\sigma_y)/2$, $v_k=\frac{k}{m^\ast}$,
$\beta_k=m^\ast[{\gamma({k^2-4{\langle k_z^2\rangle}})/{2\sqrt{3}}+\alpha eE_z}]/2$,
$\gamma_k=i\gamma k^2m^\ast/(2\sqrt{6})$, and
$\rho^l_k=\frac{1}{2\pi}\int^{2\pi}_0d\theta_{\bf   k}\rho_{\bf k}e^{-il\theta_{\bf
    k}}$. Here, $1/\tau^l_k=m^\ast N_i\int^{2\pi}_0d\theta_{\bf
  k}[1-\cos (l\theta_{\bf k})]U^2_{\bf q}/(2\pi)$ is the $l$th-order
momentum-relaxation rate.  $N_i$ stands for the bulk impurity 
density and
$U^2_{\bf q}=\int^\infty_{-\infty}\frac{dq_z}{2\pi}|\frac{e^2}{\epsilon(q)
  (q^2+q^2_z)}|^2|I(iq_z)|^2$
is the impurity scattering potential with $|{\bf 
  q}|=\sqrt{2k^2(1-\cos\theta_{\bf k})}$. $\epsilon(q)$ and the $I(iq_z)$ are the screening
function\cite{Zhou} and form factor,\cite{weng} respectively. In the strong scattering
limit, the electron  
distribution approaches isotropy in the momentum space. Therefore, one can
involve only the lowest two orders of $|l|$ ($=0$, 1) and
obtain\cite{Zhanggraphene,Zhang} 

\begin{eqnarray}
  \partial^2_x\rho^0_k=&&-2i\beta_k[\sigma_y,\partial_x\rho^0_k]+\beta^2_k[\sigma_x,[\sigma_x,\rho^0_k]]\nonumber\\ 
  &&\mbox{}+\beta^2_k[\sigma_y,[\sigma_y,\rho^0_k]].
  \label{eq5}
\end{eqnarray}
The steady-state spin vector ${\bf
  S}^0_k(x)=\mbox{Tr}[\rho^0_k(x){\bgreek\sigma}]$ then can 
be obtained from Eq.\,(\ref{eq5}) together with the boundary conditions
${\bf S}^0_k(0)={\bf S}^0_k$ and ${\bf S}^0_k(\infty)=0$. Obviously,
  ${\bf S}^0_k$ are determined by not only the spin polarization at the left
  boundary but also the electron density and temperature.
({\em i}) For the injected electrons polarized along the $x$-axis,
i.e., ${\bf S}^0_k=(S^0_k,0,0)^T$, the spin polarization
vector of the steady state is given by
\begin{eqnarray}
 \hspace{-0.3cm} {\bf P}(x)&&\hspace{-0.4cm}=\frac{1}{(2\pi)^2 N_e}\int d{\bf k} {\rm Tr}[\rho_{\bf
      k}(x){\bgreek \sigma}]\nonumber\\
  &&\hspace{-0.4cm}=\hspace{-0.1cm}\int_0^{\infty}\hspace{-0.15cm}dk\frac{k
  S^0_k}{2\pi N_e}e^{\frac{-x}{l_x(k)}}\hspace{-0.1cm}\left(\begin{array}{c}
      \sqrt{1+\Delta^2}\sin(|K_k|x+\phi)\\
      0\\
c_1 \sin(K_kx)\\
 \end{array}\right)\hspace{-0.1cm}.
  \label{eq6}
\end{eqnarray}
({\em ii}) For ${\bf S}^0_k=(0,S^0_k,0)^T$, the spin polarization vector
is
\begin{equation}
  {\bf P}(x)=\int_0^{\infty}dk\frac{k 
  S^0_k}{2\pi N_e}e^{\frac{-x}{l_y(k)}}\left(\begin{array}{c}
      0\\
      1\\
      0\\
 \end{array}\right).\\
  \label{eq7}
\end{equation}
({\em iii}) For ${\bf S}^0_k=(0,0,S^0_k)^T$, one obtains
\begin{eqnarray}
 {\bf P}(x)=\hspace{-0.1cm}\int_0^{\infty}\hspace{-0.15cm}dk\frac{-k
  S^0_k}{2\pi N_e}e^{\frac{-x}{l_z(k)}}\hspace{-0.1cm}\left(\begin{array}{c}
      c_2\sin(K_kx)\\
      0\\
      \sqrt{1+\Delta^2}\sin(|K_k|x-\phi)\\
 \end{array}\right)\hspace{-0.1cm}.
  \label{eq8}
\end{eqnarray}
Here, $\Delta={(8\sqrt{2}-11)}/{\sqrt{7}}$, $\phi=\arctan({1}/{\Delta})$,
$l_x(k)=l_z(k)=(2\sqrt{2}+1)/({\sqrt{7+14\sqrt{2}}|\beta_k|})$,
$l_y(k)={1}/{(2|\beta_k|)}$, $K_k=\sqrt{1+2\sqrt{2}}\beta_k$,
  $c_1=-\frac{4}{(1+\sqrt{2})\sqrt{1+2\sqrt{2}}}$, and
  $c_2=\frac{(20\sqrt{2}-24)\sqrt{1+2\sqrt2}}{7}$.
At low temperature, these equations
give the frequency of the spin precession in the
space domain, $K_{k_F}$, and the corresponding spin diffusion lengths
$l_x(k_F)$, $l_y(k_F)$ and 
$l_z(k_F)$ for the injected spin polarized along $x$-, 
$y$- and $z$-axes, respectively. Here, $k_F$ is the Fermi wave vector.
Since the electrons are in the quasi-equilibrium state in the strong scattering
regime, one can still treat Eq.\,(\ref{eq4}) up to the second order
and the scattering term can be approximately written as
$\rho^{\pm1}_k/\tau^{\pm1}_k$ for the inelastic scatterings. Moreover, from 
  the symmetry of the scattering matrices, $1/\tau^{0}_k=0$ and
$1/\tau^{1}_k=1/\tau^{-1}_k$ are still tenable. Therefore,
the solutions of the spin polarization are still the same as
Eqs.\,(\ref{eq6})-(\ref{eq8}). 
However, the carrier exchange effect for the electrons with different $k$ due to
the inelastic scattering mechanisms, which is not included in these equations, results in a
unique spatial precession frequency.\cite{Cheng}
As we will show latter, this effect can be clearly seen in the vicinity of the cancellation.

\section{NUMERICAL RESULTS}
To include the electron-electron and electron-phonon
scatterings, the KSBEs are solved
numerically by employing the double-side injection boundary
conditions\cite{Cheng} as 
\begin{eqnarray}
&\hspace{-0.5cm}\rho_{\bf k}(0,t)=\frac{F_{{\bf k}\uparrow}^0+F_{{\bf
    k}\downarrow}^0+(F_{{\bf k}\uparrow}^0-F_{{\bf
    k}\downarrow}^0){\bgreek\sigma\cdot{\bf \hat n}}}{2}, &({\rm for}\hspace{0.2cm} k_x>0), \label{eq9}\\
&\hspace{-2.6cm}\rho_{\bf k}(L,t)=\frac{F_{{\bf k}\uparrow}^L+F_{{\bf
    k}\downarrow}^L}{2}, &({\rm for}\hspace{0.2cm} k_x<0),\label{eq10}
\end{eqnarray}
and $\Psi(0,t)=\Psi(L,t)=0$ since no in-plane static electric field is applied.
Here, the spins are polarized along ${\bf \hat n}$. $F^{i}_{{\bf
    k}\uparrow}$ ($F^{i}_{{\bf  k}\downarrow}$), with $i=0,L$, satisfy the Fermi distribution
for the spinors parallel (antiparallel) to the ${\bf\hat
  n}$ direction. In the calculation, the Rashba coefficient $\alpha$ is
taken to be 28~\AA$^2$ (Ref.\,\onlinecite{Hassenkam}), the initial spin polarization at the
left boundary is set to be 5\% and the other parameters are taken from
Ref.\,\onlinecite{weng}. With $\rho_{\bf k}(x,+\infty)$ obtained
from the KSBEs, the spin polarization signal along 
the detection direction ${\bf \hat{m}}$
is  described by $P_m(x)=\sum_{\bf k}$Tr$[\rho_{\bf
  k}(x,+\infty)\sigma_{\bf \hat m}]/N_e(x,+\infty)$, which can be well fitted by\cite{fitting}
\begin{equation}
P_m(x)=A \exp(-x/L_{t{\bf \hat m}})\cos(K_{t{\bf \hat m}} x+\phi_t).
\label{eq11}
\end{equation}
One then defines the spin diffusion length $L_{t{\bf \hat m}}$ and the
spatial precession frequency $K_{t{\bf \hat m}}$. The
acquisition of the unique diffusion length and precession frequency from
Eq.(\ref{eq11}) reflects
the carrier exchange effect as discussed in Sec.\,II.

\begin{figure}[htb]
\includegraphics[height=9cm]{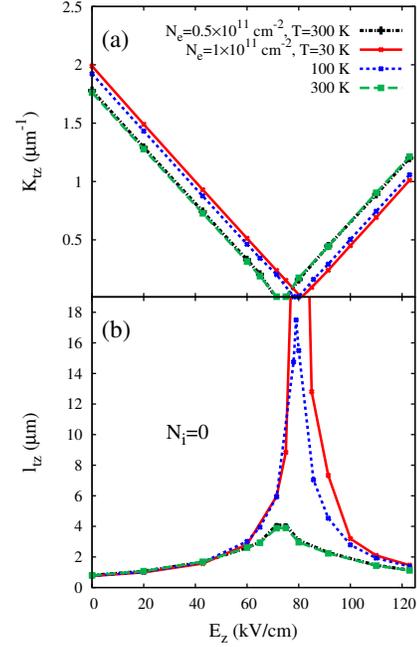}
\caption{(Color online)  $E_z$ dependence of (a) the
  spatial precession frequency and (b) the
  corresponding spin diffusion length at different temperatures and electron
  densities. The impurity density $N_i=0$ and the spin polarization is along the
  $z$-axis. }
\label{figss1}
\end{figure}

We first investigate the effect of $E_z$ by taking both injection
and detection polarization along
  $z$-direction. The  spatial precession frequency and the corresponding 
spin diffusion length as function of $E_z$ are plotted in Fig.\,\ref{figss1}.
As shown in Fig.\,\ref{figss1}(a), $K_{tz}$ is approximately
proportional to $|E_z-E_z^c|$, where the cancellation electric field $E_z^c$ is
calculated from 
Eq.\,(\ref{eq2}) ($E^c_z=82$~kV/cm at 30~K, $80$~kV/cm at 100~K, and
$77$~kV/cm at
300~K). This originates from the weak momentum dependence of
$\beta_k$ out of the strong cancellation regime, where $k^2\ll 4{\langle 
  k_z^2\rangle-2\sqrt{3}\alpha eE_z^c/\gamma}$.
In Fig\,1(b), a peak is clearly seen in the $E_z$ 
dependence of the spin diffusion length. One finds that this peak locates just
around $E_z^c$, which suggests that the peak should originate from the cancellation
effect of the Dresselhaus and Rashba terms. Furthermore,
it is revealed that the temperature and the electron density have marginal influence
on the spin diffusion length for $E_z$ far away from $E^c_z$ as
predicted by Eq.\,(\ref{eq8}). However, when $E_z$ is around
$E^c_z$, $\beta_k$ is strongly dependent on the moment and Eq.\,(\ref{eq8})
fails for the inelastic scattering as we explained in Sec.\,II. In this case, one
finds that, the spin diffusion length at the peak becomes shorter as the
temperature increases. The underlying physics lies in the fact that, for
higher temperature, the electrons disperse into a broader range in the ${\bf
  k}$ space, which means that fewer electrons are located in the regime with
the in-plane Dresselhaus term significantly canceled by the Rashba
term.\cite{Vurgaftman}
One also notices that the spin diffusion length for
$N_0=0.5\times10^{11}$~cm$^{-2}$ (with the Fermi temperature $T_F\approx 21$~K)
is almost the same as that
for $10^{11}$~cm$^{-2}$ ($T_F\approx 41$~K) at room temperature.
This is because that the inhomogeneous broadening as well as the
scattering strength (dominated by the electron-phonon scattering) is weakly 
dependent on the electron density in the non-degenerate
regime at high temperature.
When the electron gas enters into the degenerate regime, we find that the spin
diffusion length can be tuned via the electron density. For example,
the diffusion length changes from 17.5~$\mu$m to 15.2~$\mu$m as the density
increases from $10^{11}$ to $4\times10^{11}$~cm$^{-2}$ ($T_F\approx 166$~K) for
$E_z=79$~kV/cm at 100~K.

\begin{figure}[htb]
\includegraphics[height=5.4cm]{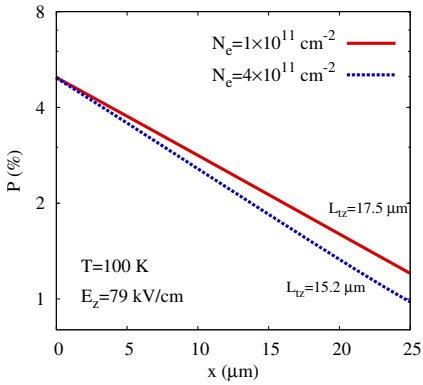}
\caption{(Color online)  The magnitude of the steady-state spin polarization as
  function of the position at 300~K with $N_0=10^{11}$~cm$^{-2}$ and
  $N_s=aN_i/N_0$. }
\label{figss2}
\end{figure}  

To observe the role of the scattering strength further, we introduce the impurities into
our GaAs QWs. The magnitude of the spin polarization
with and without impurities are plotted as function of position in
Fig.\,\ref{figss2}. In the absence of the perpendicular electric field, far away from
the cancellation condition, the additional impurity scattering channel only
slightly changes the spin diffusion length ($\sim 1$~\%) which is consistent with
Eq.\,(\ref{eq8}). However, around $E^c_z$, the spin diffusion can be
significantly influenced by the high doping density as shown in
the figure, where $L_{tz}$ changes from 4.0 (3.0) to 3.3 (2.7)~$\mu$m with
$E_z=75$~(80)~kV/cm by introducing the impurities.

\begin{figure}[htb]
\includegraphics[height=5.5cm]{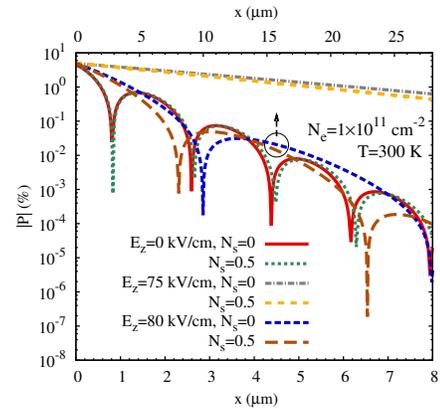}
\caption{(Color online)  The magnitude of the steady-state spin polarization as
  function of the position with different polarization directions in the absence
  of impurity. We take T=100~K and $N_0=10^{11}$~cm$^{-2}$. Here, $i$-$j$
    represents the measured spin projection along $j$-axis while the injected
    spin polarization is along $i$-axis.
}
\label{figss3}
\end{figure}

We also investigate the anisotropy of the spin
diffusion by taking the injection and detection polarizations along
different directions.
The position dependence of the magnitude of the spin polarization 
with $E_z=0$ and $79$~kV/cm are plotted in Fig.\,\ref{figss3}, where the
notation 
$i$-$j$ means the injected spin polarization
along $i$-direction and the detected one along $j$-direction.
Since the signal in the $x$-$y$, $y$-$x$,
$y$-$z$ and $z$-$y$ configurations are negligible
as predicted by Eqs.\,(\ref{eq6})-(\ref{eq8}), we only show the relevant
components in the figure. For $E_z=79$~kV/cm, in the vicinity of the
cancellation electric field, the spin polarizations for the $x$-$z$
and $z$-$x$ configurations also vanish due to the absence of 
the spatial spin precession. Interestingly, it is seen from
Fig.\,\ref{figss3} that the spin 
diffusion lengths for the $x$-$x$ and $z$-$z$ cases for $E_z=0$~kV/cm  are equal 
[$L_{tx}=L_{tz}=0.76$~$\mu$m and $L_{ty}=0.51$~$\mu$m, well consistent with the
relation $l_x=l_z\approx 1.48 l_y$ from Eqs.\,(\ref{eq6})-(\ref{eq8})], 
while the $x$-$x$ and $y$-$y$ cases share the same spin
diffusion length ($L_{tx}=L_{ty}=9.3$~$\mu$m and $L_{tz}=17.8$~$\mu$m)
for the electric field $E_z=79$~kV/cm. In the
latter case without spatial spin precession,
the drift-diffusion model performs well.\cite{Zhanggraphene} Therefore, the spin
diffusion length can be expressed by 
$L_{{\bf \hat n}}=\sqrt{D_s\tau_{s{\bf \hat n}}}$ with $\tau_{s{\bf \hat
    n}}\sim 1/\overline{(|{\bgreek\Omega}({\bf k})|^2-\Omega^2_{\bf \hat
  n})\tau_p}$ standing for the spin relaxation time. $D_s$ denotes the
spin diffusion constant and $\tau_p$ represents the momentum relaxation
time.\cite{wuReview,opticalorientation}  
According to Eq.\,(\ref{eq1}), one has $\overline{|{\bgreek\Omega}({\bf
  k})|^2-\Omega^2_{x}}=\overline{|{\bgreek\Omega}({\bf
  k})|^2-\Omega^2_{y}}\approx 3.3\overline{|{\bgreek\Omega}({\bf
  k})|^2-\Omega^2_{z}}$, which lead to the relation
$L_{x}=L_{y}\approx0.55 L_{z}$ by considering $D_s$ and $\tau_p$ free from the
small spin polarization. This agrees well with the result
  from the KSBEs ($L_{tx}=L_{ty}\approx0.52 L_{tz}$).

Finally, we should point out that the correction of the envelope function due to
the perpendicular electric field is neglected, even 
though it can modify the spin orbit coupling via the quantity $\langle k_z^2
\rangle$ and the scattering strength via the form factor in the scattering
term.\cite{Zhou,weng} This is because: ({\em i})
the modification of $\langle k_z^2 \rangle$ due to electric field is quite small
(within 3~\%); ({\em ii}) the spin diffusion is insensitive
to the scattering strength as discussed above. Direct
calculation 
shows that only slight modification of the diffusion length is introduced by
the electric field, for example, around 3~\% at 100~K with $E_z$=79~kV/cm.

\section{CONCLUSION}
In conclusion, we have investigated the spin diffusion in $n$-type (111) GaAs
QWs by solving the microscopic KSBEs. In the perpendicular electric
dependence of the spin
diffusion length, a peak due to the cancellation between the in-plane
Dresselhaus spin-orbit coupling and the Rashba term is predicted. 
We find that, for the perpendicular electric field away from the strongest
cancellation 
value, the spin diffusion length is insensitive to the electron density,
temperature, and doping density. However, in the vicinity of the
cancellation electric field, the spin diffusion length shows strong dependence
on the temperature and doping density. For high electron density, the electron
gas is degenerate, and the spin diffusion length is found to be also
affected by the electron density. Finally, we uncover the anisotropic spin
diffusion with respect to the spin polarization
direction.

\begin{acknowledgments}
We would like to thank M. W. Wu for proposing the topic as well as
directions during the investigation. One of the authors (B.Y.S) would also like
to thank P. Zhang for helpful discussions. This work was supported by the
Natural Science Foundation 
of China under Grant No. 10725417.
\end{acknowledgments}


\begin{thebibliography}{0}
\bibitem{fitting} The asymptotic standard errors are less
  than 0.1\% in the worst case.
\end{thebibliography}
\end{document}